\documentclass[10pt,twocolumn,letterpaper,groupedaddress]{revtex4}

\usepackage{multibbl}
\usepackage{graphicx,amsmath,SIunits}
\usepackage{amsmath,SIunits}
\usepackage[latin1]{inputenc}
\usepackage{latexsym,stmaryrd}
\usepackage[sort&compress]{natbib}
\usepackage{color}
\bibpunct{[}{]}{,}{n}{}{,}

\makeatletter
\newcommand*{\citenst}[2][]{%
  \begingroup
  \let\NAT@mbox=\mbox
  \let\@cite\NAT@citenum
  \let\NAT@space\NAT@spacechar
  \let\NAT@super@kern\relax
  \renewcommand\NAT@open{[}%
  \renewcommand\NAT@close{]}%
  \citep{#2}%
  \endgroup
}
\makeatother

\begin{document}

\title{Two mechanisms of disorder-induced localization in photonic-crystal waveguides}

\author{P. D. Garc\'{i}a}
\author{G. Kir\v{s}ansk\.{e}}
\author{A. Javadi}
\author{S. Stobbe}
\author{P. Lodahl}
\email{lodahl@nbi.ku.dk}
\homepage{http://www.quantum-photonics.dk/}

\affiliation{Niels Bohr Institute,\ University of Copenhagen,\ Blegdamsvej 17,\ DK-2100 Copenhagen,\ Denmark}

\date{\today}

\small

\begin{abstract}
Unintentional but unavoidable fabrication imperfections in state-of-the-art photonic-crystal waveguides lead to the spontaneous formation of Anderson-localized modes thereby limiting slow-light propagation and its potential applications.\ On the other hand, disorder-induced cavities offer an approach to cavity-quantum electrodynamics and random lasing at the nanoscale.\ The key statistical parameter governing the disorder effects is the localization length, which together with the waveguide length determines the statistical transport of light through the waveguide.\ In a disordered photonic-crystal waveguide, the localization length is highly dispersive, and therefore, by controlling the underlying lattice parameters, it is possible to tune the localization of the mode.\ In the present work, we study the localization length in a disordered photonic-crystal waveguide using numerical simulations.\ We demonstrate two different localization regimes in the dispersion diagram where the localization length is linked to the density of states and the photon effective mass, respectively.\ The two different localization regimes are identified in experiments by recording the photoluminescence from quantum dots embedded in photonic-crystal waveguides.
\end{abstract}


\maketitle

In quantum nanophotonics, low-dimensional photonic nanostructures, such as cavities or waveguides, are fabricated in order to enhance the photon-emitter interaction ~\cite{review}.\ Importantly, the photon dispersion can be engineered, enabling, e.g., slow-light transport~\cite{Baba} or efficient single-photon sources~\cite{beta-factor}.\ However, the optical properties of photonic nanostructures are often rather sensitive to unintentional but unavoidable fabrication imperfections~\cite{Gerace,Savona_optimization}.\ The interplay between order and disorder in a photonic crystal~\cite{Savona_cavities,C0} or a photonic-crystal waveguide~\cite{Vollmer,Luca,Vasco} may induce strong light confinement due to multiple light scattering.\ The underlying wave interference process leads to disorder-induced Anderson localization, which was initially developed to explain the metal-insulator phase transition for electron waves~\cite{Anderson}.\ In photonic crystals, the ability to precisely mold the dielectric medium even on a length scale smaller than the photonic-crystal unit cell implies that the photon dispersion relation can be engineered.\ Consequently, light localization processes can be modified.\ In the present work, we identify two different regimes of Anderson localization and we extract the governing localization length, $\xi$.\ In these two regimes, the localization length is linked to two different underlying properties of the photonic lattice, i.e., either the photonic density of states (DOS) or the photon effective mass.

\begin{figure}[t]
  \includegraphics[width=\columnwidth]{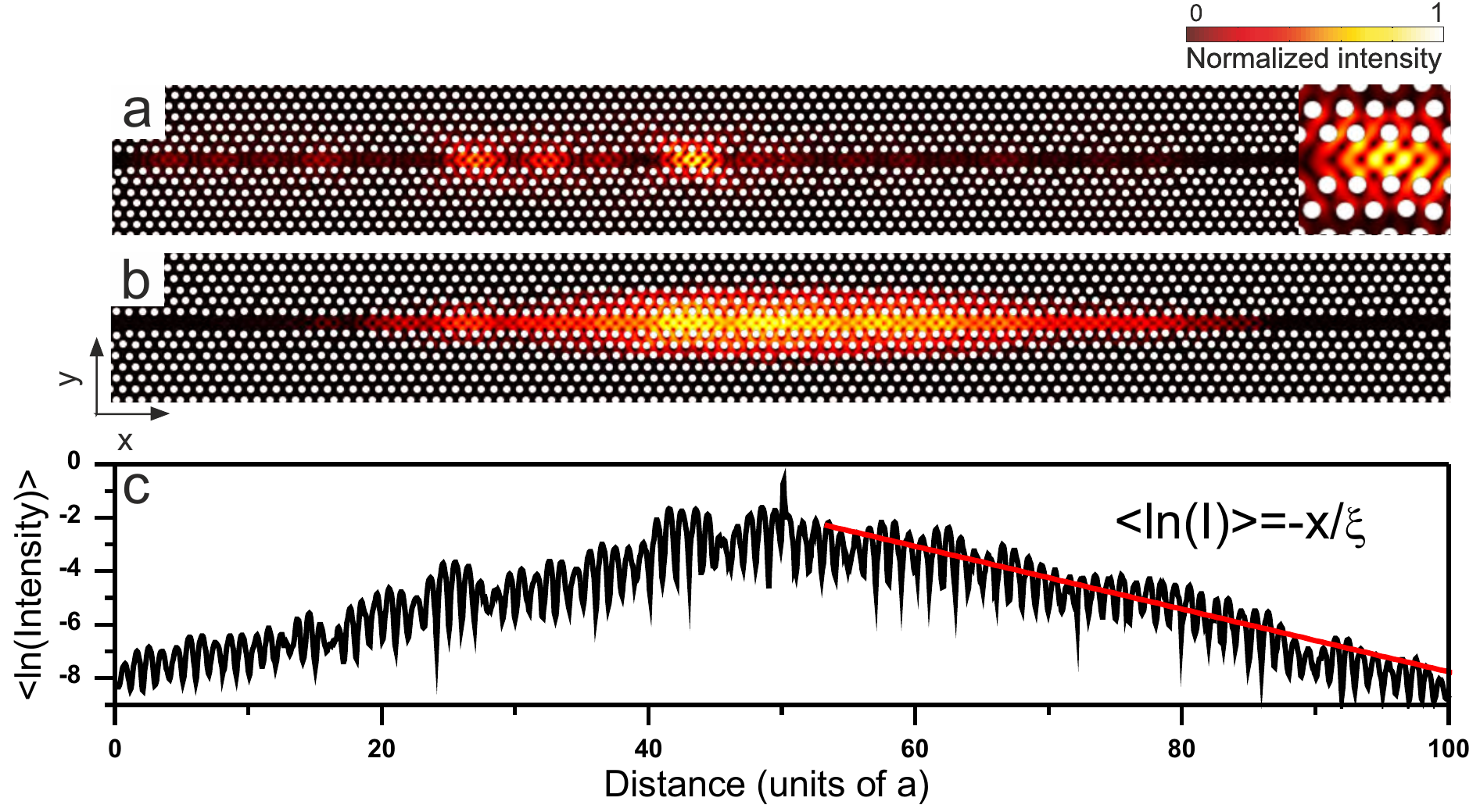}
      \caption{ \label{1} \textbf{Localization length in disordered photonic-crystal waveguides.} (a) Steady-state electromagnetic-field intensity (normalized) in a disordered photonic-crystal waveguide calculated near the cutoff frequency of the unperturbed waveguide mode.\ The parameters used for the calculation are given in the text.\ (b) Electromagnetic field intensity (normalized) emitted from a dipole source at $\omega=0.266a/\lambda$ placed at the center of a waveguide perturbed by $\sigma=0.04a$ after averaging over ten different configurations.\ (c) Ensemble-averaged electromagnetic-field intensity profile along the waveguide direction.\ The localization length can be extracted from the slope of the exponential decay.}
    \end{figure}

To reveal the two different mechanisms leading to localization, we study the scaling of the localization length, $\xi$, which is the ensemble-averaged exponential decay of the electromagnetic field intensity.\ $\xi$ is a key parameter in the localization regime determining the light transport, and therefore is related to the light-matter interaction strength between a quantum emitter and an Anderson-localized mode~\cite{Smolka2011,Tyrrestrup} as well as the emission efficiency of random lasing ~\cite{Jin}.\ To calculate $\xi$, we use two-dimensional finite-difference time-domain (FDTD) simulations~\cite{fdtd} of perturbed photonic-crystal waveguides using a freely available software package~\cite{meep}.\ We consider a hexagonal lattice of air holes forming a two-dimensional photonic crystal with a lattice constant $a=260\,\text{nm}$ and a hole radius $r=0.29a$ where a waveguide is introduced by leaving out a row of holes.\ The length of the simulation domain is $100a$, with seven rows of holes on each side of the waveguide.\ Disorder is introduced by displacing the position of the holes in the three rows on both sides of the waveguide by a random amount, $\Delta \text{\textbf{r}}$, which is normally distributed with a standard deviation $\sigma = \sqrt{\langle \Delta \text{\textbf{r}}^2\rangle}$ and $\langle \Delta \text{\textbf{r}} \rangle = 0$ where the brackets indicate the ensemble average over all configurations of random fluctuations.\ By using an effective refractive index of $n=2.76$, a $150\,\text{nm}$ thick photonic-crystal membrane can effectively be simulated in two dimensions~\cite{effective_refractive_index}, which significantly reduces the computation time.\ Inflectionless absorbers cover three lattice units at both waveguide terminations in order to mimic an open system.\ To calculate the electric-field intensity, we place a dipole source emitting at a frequency $\omega$ at the center of a waveguide perturbed by $\sigma = 0.04a$ with an harmonic time evolution.\ To compute the steady-state, we run the FDTD simulation until the time variation of the complex-field vector is the same as the light source, i.e., harmonic.\ When the holes of the waveguide are displaced randomly, imperfections lead to backscattering of the Bloch mode which creates an interference pattern along the waveguide.\ An example is plotted in Fig.~\ref{1}(a) for a frequency $\omega=0.266a/\lambda$.\ This interference pattern forms localized cavities at random positions along the waveguide and spectrally within the so-called Lifshitz tail of disordered states~\cite{Lifshitz}, which has been used to quantify the amount of disorder due to the fabrication process~\cite{quantifying}.\ After ensemble-averaging over many configurations of disorder, the intensity pattern envelope decays exponentially from the position of the source with $\xi$ as the exponential decay length of the overall envelope, cf. Figs.~\ref{1}(b) and (c).\ Note that the underlying periodic Bloch character of the mode remains after ensemble averaging, and the localization length can be extracted from the decaying envelope function.\ The exponential damping is not found in all disordered single-mode waveguide systems~\cite{Lalanne_2}, especially as the group velocity decreases.\ However, the observation of an exponential decay in photonic-crystal waveguides was predicted in Ref.~\cite{Lalanne_2}.

\begin{figure}
  \includegraphics[width=\columnwidth]{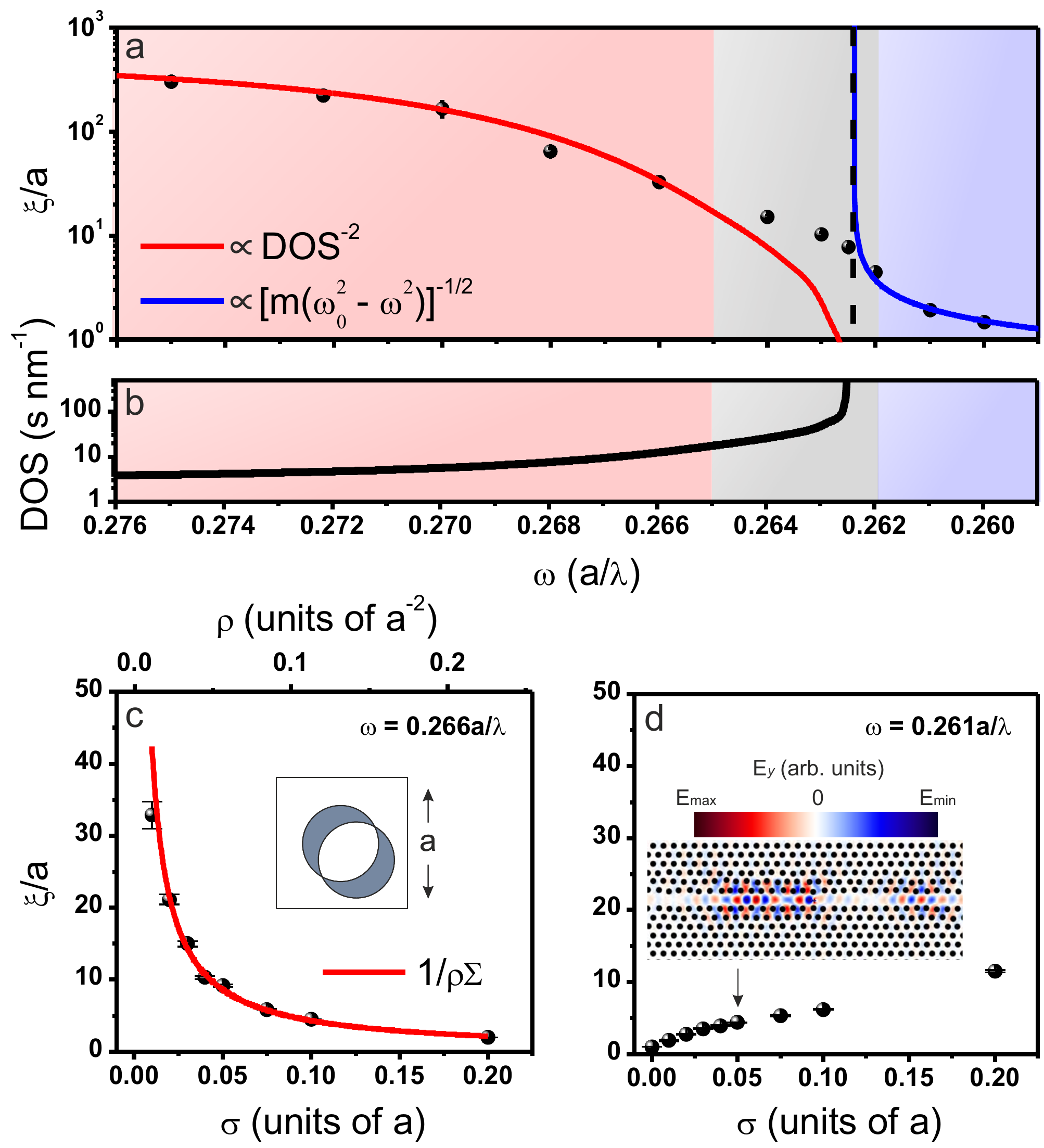}
    \caption{ \label{2} \textbf{Two regimes of disorder-induced localization} (a) Localization length calculated in a photonic-crystal waveguide with a lattice constant $a=260\,\text{nm}$ and a hole radius $0.29a$ perturbed by $\sigma=0.01a$ (black dots).\ The dashed line pinpoints the cutoff frequency of the propagating Bloch mode in the ideal structure.\ Two different scalings are observed.\ In the band region of a propagating waveguide mode, the red line shows $\xi \propto \text{DOS}^{-2}$, while the blue line is $\xi \propto \textit{m}^{-1/2}$ in the band gap region.\ (b) The black line plots the corresponding density of optical states (DOS) of a perfect photonic-crystal waveguide.\  (c) and (d) plot the localization length vs. disorder at $\omega=0.266a/\lambda$ (c, inside band) and at $\omega=0.261a/\lambda$ (d, inside band gap), respectively.\ The red line in (c) shows the scaling $\xi \propto 1/\rho \Sigma$. The inset shows the effective scattering (shaded) area when a hole is displaced from its ideal position.\ Finally, the inset in (d) shows the electric-field component perpendicular to the waveguide at $\omega=0.261a/\lambda$ and for $\sigma=0.05a$.  }
\end{figure}

Figure~\ref{2}(a) plots the dispersion of $\xi$ calculated when the holes are perturbed by $\sigma = 0.01a$.\ The black dots plot numerical data from the FDTD simulations.\ We identify two different mechanisms of localization, which correspond to two limiting situations in the band (propagating regime, red-shaded area) and in the gap (evanescent regime, blue-shaded area), respectively.\ Furthermore, a cross-over region (grey area) connects the two regimes.\ To fully understand and characterize these two localization  mechanisms, we model our numerical data with two analytical models describing the scaling of $\xi$ vs.~$\omega$.\ In the propagating regime, $\xi$ scales with the density of states as $\text{DOS}^{-2}$ (red line in Fig.~\ref{2}(a)), while in the evanescent regime this scaling is different and $\xi$ is proportional to $\textit{m}^{-1/2}$, where $\textit{m}=\left( \partial^2 \omega / \partial k^2\right)^{-1}$ is the effective photon mass in the photonic crystal, which is obtained as the inverse of the band curvature of the unperturbed mode, i.e., $\textit{m}$ is a constant associated to the band.\ We provide details of both analytical scalings later in the text.\ The described variation of $\xi$ with the scaled frequency enables controlling it and allows to tune the waveguide from operating in a propagating regime (large $\xi$) into a strongly confined regime (small $\xi$).\ Deep inside the band gap region, the mode eventually decays exponentially due to the evanescent character of light in the band gap and the attenuation length converges to the value of the Bragg length of the perfect crystal $\ell_{\text{Bragg}} \simeq a$.

The scaling of $\xi$ with the DOS in the propagating regime can be explained from a simple model as follows~\cite{Garcia2010}.\ In a single-mode waveguide $\xi \propto 1/\rho \Sigma$ ~\cite{Sheng,Enren,Beenakker,Saenz}, where $\rho$ is the density of scatterers and $\Sigma$ is the scattering cross section.\ $\Sigma$ is determined both by the excitation of the scatterer and by the subsequent scattering of light.\ The excitation of the scatterer in a photonic crystal is, in general, strongly anisotropic~\cite{PRBscattering} and, in a waveguide, it is dependent on the DOS of the incoming mode~\cite{Mcphedran}.\ The scattering process resembles that of photon emission, and the frequency scaling is well approximated by the DOS of the waveguide mode \cite{review} since the coupling to leaky modes is strongly inhibited~\cite{Hughes_purcell,Toke}.\ This leads to $\Sigma \propto \text{DOS}^2$ and therefore $\xi \propto \text{DOS}^{-2}$, and the red curve in Fig.~\ref{2}(a) displays this functional dependence.\ This is equivalent to the scaling of the localization length with the square of the group velocity, which is observed ~\cite{losses} and predicted ~\cite{Lalanne,Hughes_2005} in the literature.

\begin{figure}
  \includegraphics[width=\columnwidth]{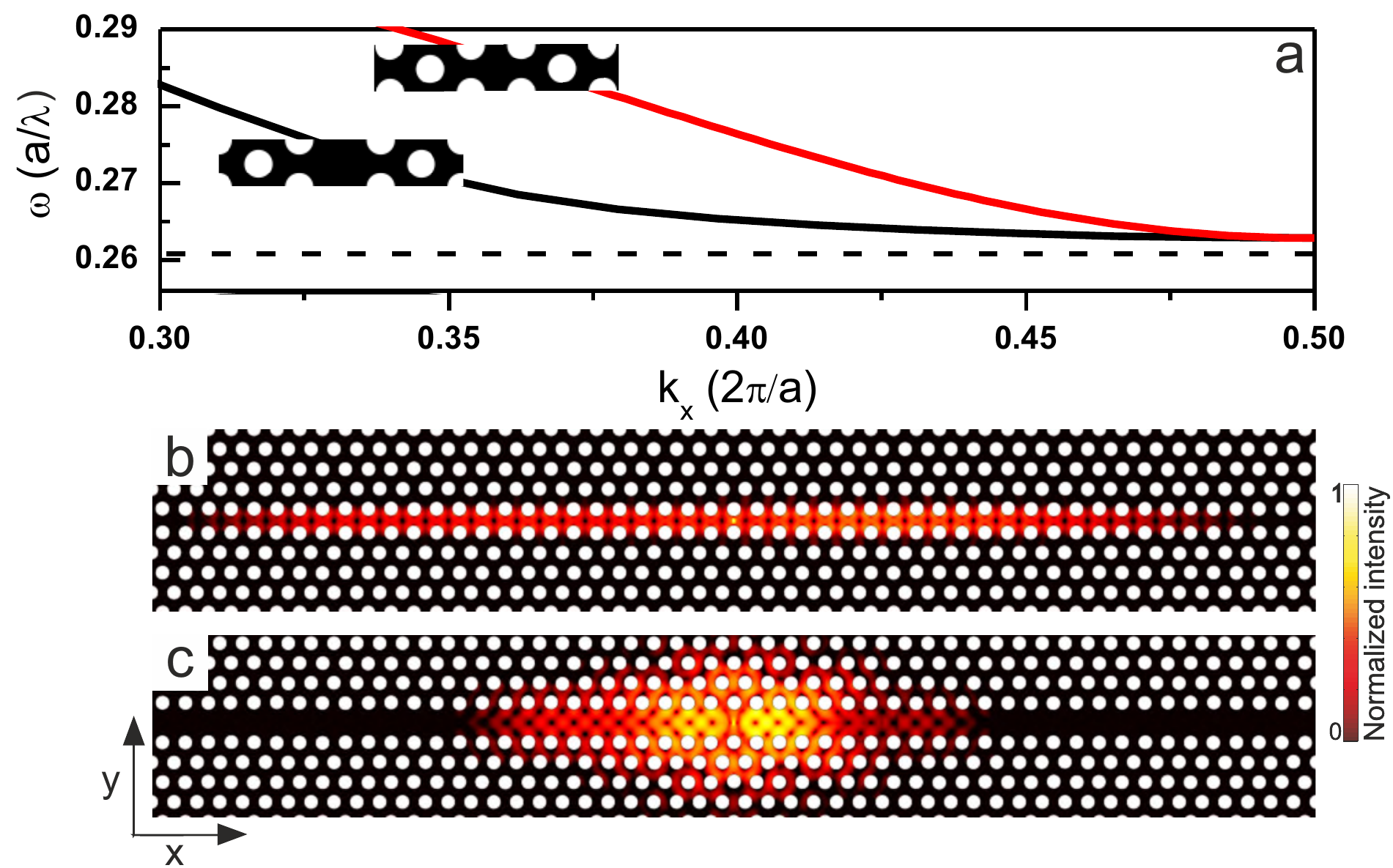}
    \caption{ \label{3} \textbf{Effective mass of photons and localization length.} (a) Dispersion relation of a photonic-crystal waveguide with $a=260\,\text{nm}$, a hole radius $0.29a$ and a standard width of $\sqrt{3}a$ (black curve) and for a stretched photonic-crystal waveguide with a reduced width of $0.60815a$ (red curve).\ By stretching the waveguide, the resulting guided mode maintains the same cutoff frequency but has an eight times larger curvature (second derivative of the dispersion curve) than in the standard case.\ (b) and (c) Calculated electric-field intensity emitted from a dipole source at $\omega=0.261a/\lambda$ (dashed line) and placed at the center of the standard stretched photonic-crystal waveguide after ensemble average over ten different configurations, respectively.}
\end{figure}

As shown in Fig.~\ref{2}(a), the scaling of $\xi$ with $\text{DOS}^{-2}$ breaks down in the cross-over regime and, particularly in the band gap, the mechanism leading to localization is different.\ Disorder perturbs the translational symmetry of the waveguide causing slight fluctuations of the cutoff frequency around the ideal value, which effectively creates barriers at random positions along the waveguide~\cite{Topolancik}.\ In the band gap, we have a set of potential barriers that confine light.\ To describe this set of random cavities, we can use the formalism developed by Slater to explain electronic transport in perturbed periodic atomic lattices~\cite{Slater} and applied to photonic lattices in Ref.~\citenst{Sargent}.\ Essentially, this formalism is the effective-mass approximation applied to photonic heterostructures which, in our case, are due to the random fluctuation of the cutoff frequency.\ This approach is also known as the \textit{envelope approximation} since it yields a solution, which results from the convolution of the ideal Bloch mode and an exponentially decaying envelope function, as shown in Fig.~\ref{1}(c).\ The frequency-dependent attenuation coefficient of this envelope function can be derived directly from Ref.~\citenst{Sargent} as $\xi/a \propto [\textit{m} (\omega_{0}^2-\omega^2)]^{-1/2}$, where $\omega$ and $\omega_{0}$ are the cut-off frequency of the perturbed and unperturbed structure, respectively, and \textit{m} is the effective photon mass.\ In the evanescent regime, i.e., when $\omega < \omega_{0}$, the perturbation creates random barriers as explained above and the dependence of $\xi$ vs.~$\omega$ is given by this effective-mass approximation, as plotted with the blue curve in Fig.~\ref{2}(a).\ A stronger confinement, i.e., a shorter $\xi$, is predicted~\cite{Vynck} for flatter bands corresponding to a larger \textit{m}.\ To illustrate this prediction, Fig.~\ref{3}(a) plots the dispersion relation in two photonic-crystal waveguides with different curvature.\ By reducing the width of the waveguide down to $0.608a$, the resulting waveguide mode (red curve) has about eight times larger curvature than the standard waveguide (black curve) while keeping the same cutoff frequency.\ The corresponding ensemble-average intensity at $\omega=0.261a/\lambda$ (dashed line in Fig.~\ref{3}(a)) is plotted in Fig.~\ref{3}(b) and Fig.~\ref{3}(c), resulting in a localization length of $\xi = 14a$ and 2a, respectively.\ This shows how \textit{heavy} photonic modes are much more sensitive to disorder than \textit{light} ones.\ The idea of an effective-mass governing localization is already pointed out in the seminal work by S. John~\cite{John}, where the Maxwell problem is approximated by an effective Schr\"{o}dinger equation in which the kinetic term is essentially determined by the curvature of the single Bloch mode under study and the potential term is given by the random potential between actual electric field modes due to disorder~\cite{Savona}.\

To further extend our analysis of the localization length, we plot $\xi$ vs.~$\sigma$ at two different $\omega$, one in the propagating regime, Fig.~\ref{2}(c), and the other one in the band-gap regime, Fig.~\ref{2}(d).\ The dependence of $\xi$ with disorder is opposite at these two frequencies: it decreases in the propagating regime while it increases in the band-gap region.\ We calculate analytically the dependence of the localization length on disorder in the propagating regime from $\xi \propto 1/\rho \Sigma$.\ For a fixed $\omega$, $\Sigma$ is also constant and $\xi$ only depends on the density of scatterers, $\rho$.\ We can, therefore, fit our numerical calculations of $\xi$ with this analytical scaling and extract a value for $\Sigma$ at this particular $\omega$.\ It tells us how strongly the imperfections in the lattice scatter light at this particular frequency.\ To do it properly, we have to quantify how much scatterer density, $\rho$, corresponds to a given amount of disorder $\sigma$.\ As we reduce here to two-dimensional calculations, $\rho$ is calculated as a scattering area per unit area ($a^2$ in our case).\ To calculate this scattering area, we assume that only deviations from the ideal hole position scatter light~\cite{Koenderink}.\ As shown in the inset of Fig.~\ref{2}(c), this scattering (shaded) area can be expressed in terms of the intersection between the perturbed and unperturbed hole as $2[ \pi \text{r}^2 - 2 \text{r}^2 \arccos(| \Delta  \text{\textbf{r}} |/2\text{r}) + | \Delta  \text{\textbf{r}} | \sqrt{\text{r}^2 - (| \Delta  \text{\textbf{r}} |/2)^2} ]$, where r is the hole radius and $| \Delta  \text{\textbf{r}} | = \sigma$ is the hole random average displacement.\ This is how we obtain the relation between $\rho$ and $\sigma$ which allows us to estimate the scattering cross section as $\Sigma = (2.02 \pm 0.04)a = (6.9 \pm 0.9)r$ at $\omega = 0.266 a/ \lambda$, which is the cross section of a \textit{full hole} and shows a dramatically enhanced scattering response.\ To put these results into perspective, the behavior observed here in the propagating regime is in agreement with previous full-Bloch mode analysis in disordered waveguides~\cite{Lalanne}, which have been focused on the evolution of $\xi$ with the group index, i.e., in the propagating regime.\ In contrast, experiments with embedded quantum emitters in disordered photonic-crystal waveguides~\cite{Smolka2011} reported the increase of $\xi$ with disorder. The latter is explained by the fact that the experiment was spectrally averaged over the full frequency range where localized modes were observed, i.e., also band-gap localized modes were included.

\begin{figure}
  \includegraphics[width=\columnwidth]{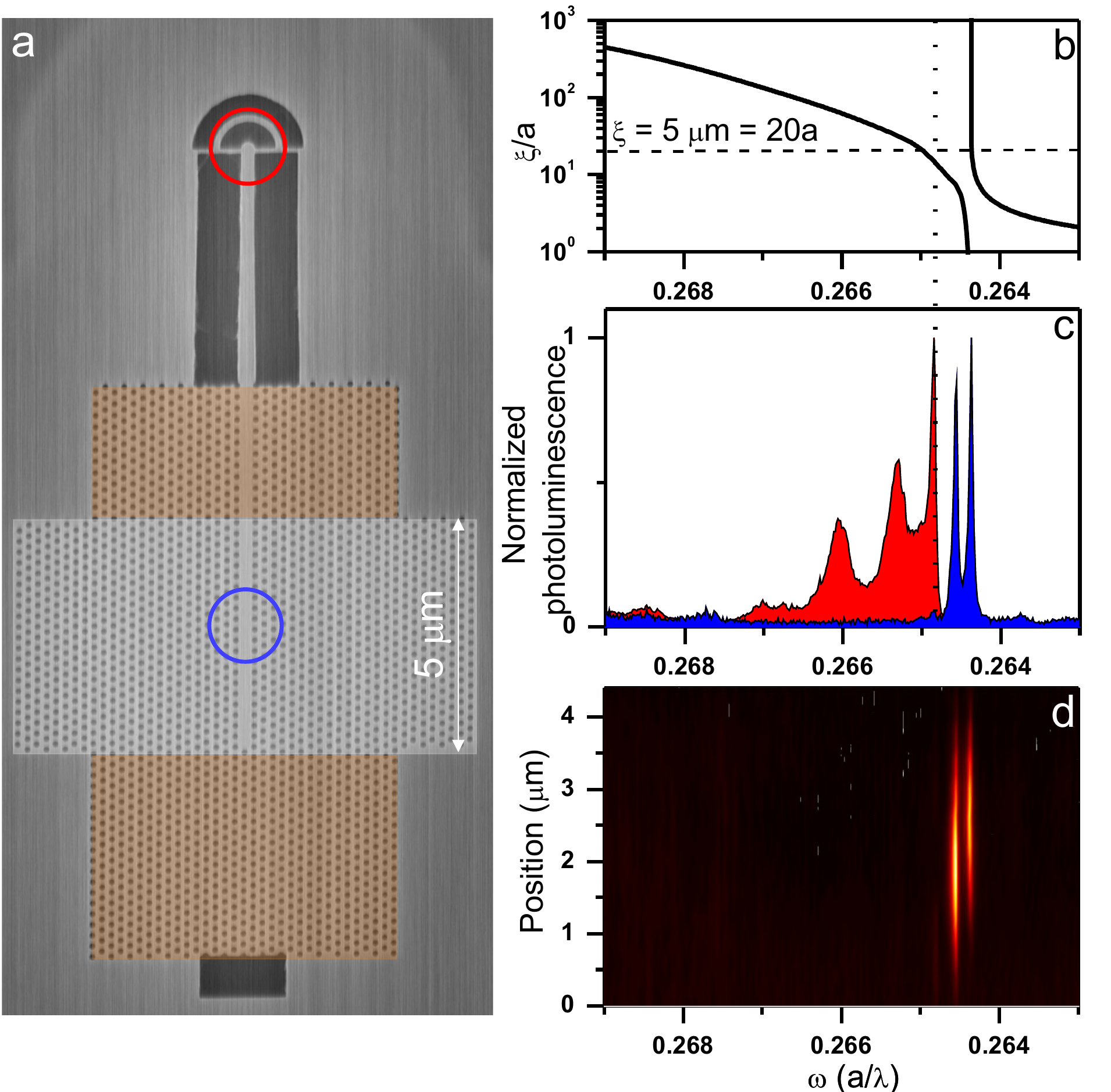}
    \caption{ \label{4} \textbf{Experimental extension of localized modes.} (a) Scanning-electron micrograph of a photonic-crystal waveguide with a $5\,\micro\text{m}$-long slow light section (white-shaded region) and parameters $a=252\,\text{nm}$ and $r=0.27a$.\ The sample contains InGaAs quantum dots which are excited in the slow-light section (blue circle) and their photoluminescence is collected either from the same spot or from the circular grating (red circle) after propagation through a fast-light section corresponding to $a=270\,\text{nm}$ and $r=0.27a$ (orange-shaded region).\ (b) The localization length calculated for the parameters of the fabricated slow-light section of the waveguide, where scattering is most pronounced.\ The dashed line marks the localization length corresponding to the length of the slow-light region of the sample.\ (c) High-power photoluminescence spectra probing light from the quantum dots that are coupled to the waveguide mode and subsequently coupled out through the grating (red-filled line) or localized in the slow-light section of the waveguide and coupled out through leaky modes (blue-filled line).\ (d) Normalized photoluminescence intensity of the localized modes measured while raster scanning the sample under high-excitation power.}
\end{figure}

Near-field scanning experiments have recently measured~\cite{Vynck} localized modes in the band-gap region of state-of-the-art photonic-crystal waveguides affected only by a small amount of fabrication disorder, i.e., $\sigma = 0.002a$.\ In this case the localized modes should be determined by the effective mass.\ To probe experimentally the localized modes in the band-gap region, we have designed very short photonic-crystal waveguides in order to discriminate them from localized modes appearing in the propagating regime that were already studied in detail elsewhere~\cite{Garcia2010}.\ Optically pumped quantum dots embedded in the waveguides efficiently excite the localized modes~\cite{Light-matter}.\ To reveal the relation between $\xi$ and the effective mass in the evanescent regime, we fabricate waveguides with different engineered dispersion relation.\ In detail, our samples consist of a $20-\,\text{nm}$-thick GaAs layer sandwiched between $15-\,\text{nm}$- and $35-\,\text{nm}$-thick AlGaAs barriers and constituting a $\emph{p}-\emph{i}-\emph{n}$ diode.\ A layer of high-density ($\sim 100\,\micro\text{m}^{-2}$) self-assembled InGaAs quantum dots is embedded in the intrinsic GaAs layer.\ A mesa structure is defined by standard optical lithography and wet etching, and connected through the bottom $\emph{n}$-type and top $\emph{p}$-type ohmic contacts.\ The sample biased at a constant electric field of around $100\,\text{kV}/\text{cm}$~\cite{Midolo2015}.\ The central part of the nanostructure consists of a so-called slow-light waveguide section, the white-shaded region in Fig.~\ref{4}(a), designed to have a high group index.\ A fast-light section with a low group index, the orange-shaded region in Fig.~\ref{4}(a), is used to efficiently extract light from the slow-light section out through a circular Bragg grating.\ The sample is terminated at the opposite side by a photonic crystal which, together with the grating, forms a weak cavity giving rise to Fabry-Perot resonances.\ For optical measurements, the sample is placed in a liquid-helium flow cryostat and cooled down to $10\,\text{K}$.\ A continuous-wave Ti:sapphire-laser beam tuned to a wavelength of $883\,\text{nm}$ is focused to a $1.5\,\micro\text{m}$ spot through a microscope objective of $\textrm{NA} = 0.6$ and excites the slow-light section of the waveguide from the top (details can be found in Ref.~\cite{Luca}).\ The photoluminescence from the quantum dots is collected either from the excitation spot or from the circular grating through a single mode fiber, cf.~Fig.~\ref{4}(a).

Figure~\ref{4}(b) plots the calculated localization length for the slow-light section assuming a fabrication disorder of $\sigma = 0.005a$ ~\cite{quantifying}.\ The total length of the slow-light region of the waveguide sets the crossover to Anderson localization at $\xi = 5\,\micro\text{m} = 20a$, where $a=252\,\text{nm}$ in this case.\ At frequencies below this onset, marked with a vertical dotted line in Fig.~\ref{4}(b), light emitted by the quantum dots in the slow-light region couples to the waveguide and propagates out of the structure.\ The red curve in Fig.~\ref{4}(c) plots the photoluminescence collected from the grating which shows a clear cutoff corresponding to the crossover.\ Increasing the frequency, light becomes effectively localized giving rise to sharp resonances in the photoluminescence spectrum collected directly from the waveguide, as plotted by the blue curve in Fig.~\ref{4}(c).\ Figure~\ref{4}(d) shows the photoluminescence measured after raster-scanning along the waveguide with a step size of $200\,\text{nm}$.\ From this type of spatial scan, we extract the mode spatial extension length - $\ell$ - as twice the distance over which the normalized intensity is reduced to 1/e.\ We note that $\ell$ only becomes the localization length $\xi$ after ensemble averaging over configurations of disorder.\ However, for the strongly confined modes investigated here even a single realization constitutes a useful estimate of the statistical extent of the mode.\ Increasing the hole radius of the waveguide while keeping fixed $\textit{a}$ results in a flatter dispersion relation, i.e., a larger effective mass.\ For waveguides with \textit{r/a} equal to $0.270$, $0.277,$ and $0.285$, respectively and the same lattice constant, we extract the values  $2.4\,\micro\text{m}$, $1.7\,\micro\text{m}$ and $1.4\,\micro\text{m}$ for $\ell$.\ This monotonic decrease of $\ell$ versus \textit{m} is experimental evidence of the predicted light-localization dependence on the effective mass.

In conclusion, we present a combined numerical and analytical analysis of Anderson localization in photonic-crystal waveguides.\ Our numerical simulations reveal two different mechanisms leading to localization in these structures.\ In the propagating regime, the backscattering of the perturbed Bloch mode leads to the random interference pattern.\ Inside the band gap, random fluctuations of the cutoff frequency is the mechanism behind localization.\ In the propagating regime, $\xi$ is governed by the density of optical states, in the band-gap regime $\xi$ is determined by the effective mass of the photonic band.\ These two mechanisms lead to opposite dependencies on disorder.\ Understanding the different mechanisms leading to localization is crucial to design structures which are more robust against disorder as, for example, by increasing the curvature of the Bloch mode.\ In addition, as the localization length is determining the average mode volume of localized modes~\cite{Jin}, our analysis also shows the path to exploit disorder-induced cavities optimally for light-matter interaction~\cite{Light-matter}.

\textbf{Acknowledgements}

We gratefully acknowledge financial support from the European Research Council (ERC Advanced Grant "SCALE", project ID 669758), Danish National Research Foundation (Centre of excellence "Hybrid Quantum Networks"), The Villum Foundation, and the Danish Council for Independent Research.


\end{document}